\documentclass[fleqn,prd,superscriptaddress,showpacs,nofootinbib,preprintnumbers]{revtex4}

\usepackage{amsmath}
\usepackage{amsfonts}
\usepackage{graphicx}
\usepackage{dcolumn}
\usepackage{fancyhdr}
\usepackage[mathscr]{euscript}
\def\hs{\qquad} 
\def\beq{\begin{eqnarray}} 
\def\eeq{\end{eqnarray}} 
\def\at{\left(} 
\def\aq{\left[} 
\def\ag{\left\{} 
\def\cp{\right.} 
\def\ct{\right)} 
\def\cq{\right]} 
\def\lap{\Delta\,} 
\def\ii{\infty}
\def\segue{\qquad\Longrightarrow\qquad} 
\def\al{\alpha}
\def\be{\beta}

\def\ka{\kappa}

\def\si{\sigma}

\def\La{\Lambda}
\def\Si{\Sigma}
\def\Om{\Omega}

\newcommand{\bea}{\begin{eqnarray}}
\newcommand{\eea}{\end{eqnarray}}
\newcommand{\beaa}{\begin{eqnarray*}}
\newcommand{\eeaa}{\end{eqnarray*}}

\def\nn{\nonumber}

\begin{document}

\title{\bf Einstein gravity with Gauss-Bonnet entropic corrections}

\author{
 Guido Cognola\footnote{cognola@science.unitn.it}}
\affiliation{Dipartimento di Fisica, Universit\`a di Trento \\
and Istituto Nazionale di Fisica Nucleare \\
Gruppo Collegato di Trento, Italia}

\author{Ratbay Myrzakulov\footnote{rmyrzakulov@gmail.com}}
\affiliation{Eurasian International Center for Theoretical Physics\\ 
and Department of General and Theoretical Physics, \\
Eurasian National University, Astana 010008, Kazakhstan}

\author{Lorenzo Sebastiani\footnote{l.sebastiani@science.unitn.it}}
\affiliation{Eurasian International Center for Theoretical Physics\\ 
and Department of General and Theoretical Physics, \\
Eurasian National University, Astana 010008, Kazakhstan}

\author{Sergio Zerbini\footnote{zerbini@science.unitn.it}}
\affiliation{Dipartimento di Fisica, Universit\`a di Trento \\
and Istituto Nazionale di Fisica Nucleare \\
Gruppo Collegato di Trento, Italia}

\begin{abstract}
A toy model of Einstein gravity with a Gauss-Bonnet classically ``entropic'' 
term mimicking a quantum correction is considered. 
The static black hole solution due to Tomozawa is found and generalized with 
the inclusion of non trivial horizon topology, 
and its entropy evaluated deriving the first law by equations of motion. 
As a result the Bekenstein-Hawking 
area law turns to be corrected by a logarithmic area term. 
A Misner-Sharp expression for the mass of black hole is found. 
Within a Friedmann-Lema\^itre-Robertson-Walker (FLRW) cosmological setting, 
the model is used in order to derive modified Friedmann equations. 
Such new equations are shown to reproduce the first law with the same 
formal entropy and quasi local energy of the static case, but here 
within a FLRW space-time interpreted as a dynamical cosmological black hole.
A detailed analysis of cosmological solutions is presented, and it is shown that the
presence of the correction term provides regular solutions and interesting phases of acceleration and decelerations,
as well as, with negligible matter, exact de Sitter solutions.

\end{abstract}
\maketitle 


\section{Introduction}

In a recent paper \cite{tomo12}, Tomozawa has put forward an argument 
to deal with finite one-loop quantum corrections to Einstein gravity,
by reconsidering an old result of himself. 
The techniques within his proposal refer to are well known and can be found in the celebrated 
texbook \cite{BD}. 
Here, we present a simplified approach motivated by alternative arguments, 
and in some sense, also inspired by the entropic 
approach proposed in Refs.~\cite{smoot1,smoot2}, 
where Lagrangian surface terms are taken into accout.

The paper is organized as follows.
In Section II, we will give a short review of the traditional semiclassical approach to quantum corrections of Einstein's gravity.
In Section III, the ``entropic'' approach is introduced and 
the spherically symmetric static case is investigated. 
In Section IV, the model is extended to the spherically symmetric 
dynamical FLRW space-time
and in Section V some explicit solutions are presented. Conclusions are given in Section VI.

\section{Brief review of quantum corrections to gravity}

Before discussing the model which mimics the Tomosawa proposal, 
we review the more traditional treatment as proposed by
Starobinsky in the seminal paper \cite{staro}. 
The argument is based on the so called semiclassical gravity approach, 
where the backreaction of quantum fields are taken into account, 
in order to correct the classical Einstein equation, namely
\beq
G_{ij}=R_{ij}-\frac{1}{2}g_{ij}\,R =\chi\langle T_{ij}\rangle\,,\hs\hs \chi=8\pi G_N\,, 
\eeq
$g_{ij}$, $R$ and $R_{ij}$ being respectively the metric, the scalar curvature 
and Ricci tensor, $G_N$ the Newton constant, 
and $\langle T_{ij}\rangle$ the vacuum expectation value 
of quantum energy stress tensor, which in general is not explicity known. 
As usual, units measure are chosen in such a way that the speed of light is equal to one.

By taking the trace of equation above one has
\beq
R=-\chi\langle T_{i}^i\rangle\,,
\label{adte}\eeq
in which the stress tensor trace appears. 
When one is dealing with a conformally coupled quantum field, 
a quantum conformal anomaly is present (see the review paper \cite{duff}). 
In four dimension it reads
\beq
\langle T_{i}^i\rangle=-(\alpha W+\beta G+\delta \lap R)\,.
\label{t}
\eeq
Here $W$ is the ``square'' of Weyl tensor $C_{ijrs}$ and $G$ 
the Gauss-Bonnet topological invariant. They read
\beq
W=C^{ijrs}C_{ijrs}\,,\hs\hs G=R^{ijkl}R_{ijkl}-4R^{ij}R_{ij}+R^2\,. 
\eeq
The coefficients $\alpha$, $\beta$, and $\delta$ 
depend on the number of conformal fields present in the theory. 
In some conformal field theories \cite{HS}, 
one has $\delta=0$ and $\alpha=\beta=\frac{N^2}{64\pi^2}$, 
$N$ being a very large parameter. 

One might try to solve the anomaly driven trace equation (\ref{adte})
within the static spherically symmetric (SSS) Ansatz where metric has the form
\begin{equation}
ds^2=-a(r)dt^2+\frac{d r^2}{a(r)}+r^2\,d\si_2^2\,.
\label{metric0}
\end{equation}
The account of conformal anomaly contribution to spherically-symmetric 
space-time solutions (SdS BH or wormholes) has been done in Ref.~\cite{OdiBH} (see also Ref.~\cite{Davood} for the case of extended theories of gravity).
In such a case one has
\beq
R=-\frac{1}{r^2}\,\frac{d^2}{dr^2}\,\aq r^2(a-1)\cq\,,\hs
G=\frac{2}{r^2}\,\frac{d^2}{dr^2}\,\aq a-1\cq\,,\hs
W=\frac{r^2}{3}\,\aq\frac{d^2}{dr^2}\,\at\frac{a-1}r\ct\cq^2\,.
\eeq
As a result, the quantum corrected Einstein trace equation reduces to
\beq
\frac{d^2}{dr^2}\,\aq r^2(a-1)+2\chi\be\,(a-1)^2\cq
+\frac{\chi\al}{3}\,\aq\,r^2\,\frac{d^2}{dr^2}\,\at\frac{a-1}r\ct\cq^2=0\,,
\label{at} 
\eeq
This is a second order non linear equation in the unknown variable $a(r)$.
Its exact solution appears to be difficult to find.
However, one can directly verify that there exists the de Sitter solution
\beq
a(r)=1-\frac{r^2}{2\chi\beta}\,,
\eeq
which correponds to the famous solution found by Starobinsky in FLRW coordinates.

On the other hand, putting $\xi=\chi\alpha=\chi\beta$, one can find
perturbative solutions assuming $\xi$ to be a small quantity, that is 
\beq
a(r)=a_0(r)+\xi\,a_1(r)+\xi^2\,a_2(r)+...
\eeq
Starting from $\xi=0$ one finds the unperturbed solution 
\beq
\frac{d^2}{dr^2}\aq r^2(a-1)\cq=0\segue a(r)\sim a_0(r)=1-\frac{c_1}{r}-\frac{c_2}{r^2}\,,
\label{at0} 
\eeq
$c_1,c_2$ being constants of integration. 
If we choose $c_1=2M$ and $c_2=0$, then such a solution may be interpreted 
as the Schwarzschild solution generated by a body of mass equal to $M$.

Now, by taking into account of first order perturbation we get
\beq
a(r)\sim a_0(r)+\xi\,a_1(r)=1-\frac{c_1+\xi c_3}{r}
 -\frac{c_2+\xi c_4}{r^2}-\frac{4\xi c_1^2}{r^4}
 -\frac{8\xi c_1c_2}{r^5}-\frac{22\xi c_2^2}{5r^6}\,,
\eeq
$c_3,c_4$ being further constants of integration. In principle, one may investigate this quantum corrected approximate static solution, 
but this will be not done here. 

In Ref.~\cite{cai} (see also \cite{son}), an attempt has been made to solve the anomaly driven trace equation, but 
in the special case of negligible $\al$. In fact, for $\al=0$ and $\xi=k\be$ one obtains
\beq 
a(r)=
1-\frac{r^2}{4\xi}\aq1\pm\,\sqrt{1+\frac{16\xi^2}{r^4}\,\at1+\frac{c_1}{\xi}+c_2r\ct}\cq\,,
\eeq
where $c_1,c_2$ are arbirary constants. From the mathematical point of view 
this is a quite interesting result, because it might provide quantum 
corrections to the Schwarzschild solution, but
unfortunately it is not physically relevant since 
$\alpha$ and $\beta$ are quantities of the same order. For this reason
such an approach does not seem completely adequate to describe
physical situations.

\section{Entropic corrected static spherically symmetric metric} 
Let us come back to Tomosawa proposal, reformulating it within a classical Lagrangian approach and working first in $n$ dimensions, and then making a ``entropic'' dimensional reduction to $n=4$, also in the spirit of other different approaches (see, for example, \cite{smoot1,smoot2,padma}).

It is well known that in a four dimensional manifold the Gauss-Bonnet term does not contribute 
to the classical equations of motion. 
At quantum level, the situation changes due to the regularization procedure, 
and this is the key observation made in Ref.\cite{tomo12}. 
In order to mimick quantum corrections due to Gauss-Bonnet invariant or alternatively 
in order to activate such a term in the variational principle, 
we will make use of the fact that the functional variation of the classical 
action in arbitrary $n$ dimensions does not commute with the limit $n\to4$. 

Thus, let us consider the following classical action 
\beq
I=\frac{1}{2\chi}\int d^nx\sqrt{-g}\left(R-2\Lambda+\frac{\xi\,G}{(n-4)}\right)\,,
\label{It}
\eeq
where $\La$ is the cosmological constant, $G$ the Gauss-Bonnet invariant in $n$ dimensions
and $\xi$ an arbitrary parameter, which will be
assumed to be proportional to $\chi\beta$.
We shall derive the field equations and at the end of calculation we shall perform the 
$n\to4$ limit.

We shall look for static, spherically symmetric solutions (SSS)
with arbitrary horizon topologies. The generic metric reads
\begin{equation}
ds^2=-a(r)b^2(r)dt^2+\frac{d r^2}{a(r)}+r^2\,d\Si_k^2\,,
\label{metric1}\end{equation}
where $a,b$ are arbitrary functions to be determined and
$d\Si_k^2$ is the metric of a maximally symmetric $n-2$ dimensional manifold,
which will reduce to the metric $d\si_k^2$ of a two dimensional 
maximally symmetric manifold in the limit $n\to4$, that is
\beq
d\Si_k^2 \to d\si_k^2=\frac{d\rho^2}{1-k\rho^2}+\rho^2 d\phi^2\,,\hs\mbox{ for }n\to4\,.
\nn\eeq
Here, $k=1,0,-1$ respectively for spherical, toroidal and hyperbolic topology horizons.

%
%

Within this static Ansatz, the Lagrangian becomes a function of 
$a[r],b[r]$ and their derivatives, that is
\beq
I=\Om\,\int dr\,r^{n-1}\,L(a,a',a'',b,b',b'')\,,
\label{LI}\eeq
where $\Om$ is a constant factor due to the integration on all variable except $r$.

Now, by means of the Weyl's method discussed for example in Ref. \cite{seba}, 
one obtains the field equations, which in the limit $n\to4$ are regular and read 
\beq
\frac{b'}{rb}\,\left(1+2\xi\,\frac{k-a}{r}\right)=0\,,
\label{z1}\eeq
\beq
\frac{1}{r^2b}\,\frac{d}{dr}\,\aq (k-a)r-\frac{\Lambda r^3}3+\frac{\xi(k-a)^2}r\cq=0\,.
\label{z}
\eeq
These equations can be obtained by an effective classical Lagrangian,
which mimics quantum corrections, obtained as 
the limit $n\to4$ of the one in (\ref{LI}) (apart total derivatives). 

Equation (\ref{z1}) has the trivial solution $b=\text{Const}$, 
and we can choose $b=1$ without loosing generality,
while equation (\ref{z}) gives
\beq
a(r)=k+\frac{r^2}{2\xi}\left(1\pm\sqrt{1+4\xi\left(\frac{C}{r^3}
 +\frac{\Lambda}{3}\right)}\right)\,,
\label{tt}
\eeq
where $C$ is a constant of integration. Equation (\ref{tt}) represents the topological 
generalization of Tomozawa black hole solution discussed in \cite{tomo12} 
in the presence of a non vanishing cosmological constant~\cite{supercai}. 
It is the anologue of black hole solution of Lovelock gravity 
in five dimensions \cite{deser1,h}. 
Furthermore, it is formally identical to the black hole solution 
found in \cite{KS}, as a particular limit of 
Horawa-Lifsits gravity considered in \cite{hora}. 

What about the meaning of integration constant $C$? 
A possible approach consists in discussing the limit 
$\xi\to0$ of solution (\ref{tt}) as in \cite{tomo12}. 
The limit is finite only for the solution in (\ref{tt}) 
with the minus sign in front of the square-root. In such a case the result is
\beq
a(r)=k-\frac{C}{r}-\frac{\Lambda r^2}{3}\,. 
\eeq
A a consequence one may conclude that $C=2M$, $M$ being the mass of black hole. 

Another approach consists in the investigation of the Clausius 
relation $dM=T_HdS_H$, which relates mass, temperature and entropy associated with the 
given black hole solution, as a direct consequence of field equations.
As we shall see, this approach has the merit to give informations 
about the entropy $S_H$ of the black hole solution. 

First let us study under which conditions the solution (\ref{tt}) represents a black hole. 
As it is well known, one should have real positive solution of $a(r_H)=0$, namely
\beq
\xi k^2-Cr_H+kr_H^2-\frac{\Lambda}{3}r_H^4=0\,.
\label{c}
\eeq
The general solution of this quartic algebraic equation is quite complicated 
and we will not write down its explicit form. 
For our purposes it is sufficient to know that, depending on parameters,
there exist real positive roots.
For example this is certainly true in an asymptotically flat manifold, 
that is for $\Lambda=0$ and $k=1$. In such a case one gets 
\beq
r_H=\frac{C}{2}\left( 1+\sqrt{1-\frac{4 \xi }{ C^2}}\right)\,,
\eeq
which is positive for $\xi<C^2/4$ and so it defines an event horizon.

Associated with any event horizon there exists a Hawking temperature
given by
\beq
T_H=\frac{\ka_H}{2\pi}\,, \hs\hs\ka_H=\frac{a'_H}{2}\segue a'_H=4\pi T_H\,,
\eeq
$\ka_H$ being the surface gravity related to the horizon at $r_H$.
This result is robust and can be derived by several alternative 
methods \cite{haw,PW,ang}. 

Following \cite{gorbu} we evaluate the equation of motion (\ref{z}) at the horizon, 
where $a_H=0$ and $a'_H=4\pi T_H$. We obtain
\beq
4\pi T_H\,\left(r_H+\frac{2k\xi}{r_H}\right)=k-\Lambda r_H^2-\frac{\xi k^2}{r_H^2}=
\frac{dC}{dr_H}\,,
\eeq
where the latter expression is a direct consequence of (\ref{c}), obtained by assuming
$\La$ to be a given parameter and the integration constant $C=C(r_H)$ to depend only on 
the horizon radius.

Now, introducing the horizon area $A_H=V_kr_H^2$, $V_k$ being the measure of the 
unit surface (for example $V_1=4\pi$) we can write the latter equation in 
the ``thermodynamical'' form
\beq
d\at\frac{C}{2}\ct=T_H\,d\left(\frac{\pi A_H}{V_k}+2\pi k\xi\,\ln\frac{A_H}{V_k}\right)\,.
\eeq
It is quite natural to interpret this identity as the Clausius relation 
$T_HdS_H=dE_H$ for the black hole solution with entropy $S_H$ 
and quasi-local energy $E_H$ evaluated on the horizon given respectively by
\beq
S_H=\frac{\pi A_H}{V_kG_N}+2\pi k\xi\,\ln\frac{A_H}{V_kG_N}\,,
\label{s}\eeq
and by 
\beq
E_H=\frac{C}{2G_N}=\frac{1}{2G_N}\left(kr_H-\frac{\Lambda}{3}r_H^3+\frac{\xi k^2}{r_H}\right)\,.
\label{c1}\eeq
This result gives the Energy of the black hole as classical Misner-Sharp 
mass plus a correction which depends on the parameter $\xi$. 
Furthermore, with regard to the corrected black hole entropy, 
the area law  ($S_H=\pi A_H/V_k G_N$) does not hold, because a quantum logarithmic correction is present. 
This is a well known general result, and it has been derived many times.
It has been proposed in \cite{cai}, and it appears in the
quantum field theory treatment of black hole entropy 
with heat kernel techniques \cite{dima,mann}, or loop gravity \cite{kaul} 
or other approaches (see the recent paper \cite{sen} and the references 
therein). 

It should be noted that for toroidal black hole, 
present when $\Lambda<0$, the correction is absent while for 
spherical and hyperbolic black holes the corrections have opposite sign. 

The expressions (\ref{c1}) and (\ref{s}) are depending on quantities like $r_H,A_H$, 
and $\chi_H$ which are scalars for a generic spherically symmetric (dynamical) space-times. 
It also follows that Clausius relation should 
have this covariance form. We will verify this fact in a dinamical 
spherical symmetric space-time as the FLRW one.

\section{Entropic corrected FLRW space-time} 

Anomaly driven FLRW models have been already considered 
in the past \cite{staro,vilenkin,hu,hu2,hertog}. 
Here we shall consider the toy model described by the action (\ref{It}), but 
in $n$-dimensional, spatially flat FLRW space-time defined by means of the metric
\beq
ds^2=-e^{2\eta(t)}dt^2+a(t)^2 (dx_1^2+...dx_{n-1}^2)\,,
\eeq
where $\eta(t),a(t)$ are arbritary functions, which will play the role of Lagrangian coordinates.
The quantity $\eta(t)$ will be set equal to zero at the end of calculations. 
With this choice the parameter $t$ will become the standard cosmological time.

In the total action we must also include classical matter 
described by a perfect fluid with density $\rho(t)$ and a pressure 
$p=w\rho(t)$, $0\leq w\leq\frac{1}{3}$ being a constant.
Thus the total classical action will read 
\beq
I=\frac{1}{2\chi}\int d^nx\,\sqrt{-g}\,\left(R-2\Lambda-\frac{\xi\,G}{(n-4)}\right)+I_m\,,
\label{Itm}
\eeq
$I_m$ being the action of matter which assumes the form
\beq
I_m=-\frac{1}{2}\int d^nx\,\sqrt{-g}\,g^{ij}\,\aq(\rho+p)u_iu_j+(\rho-p)g_{ij}\cq\,.
\eeq
Making the variations with respect to $\eta(t)$ and $a(t)$ in arbitrary dimension $n$, then
taking the limit $n\to4$ and finally putting $\eta(t)=0$ we obtain the 
two equations 
\beq
H^2=\frac{\chi\rho}3+\frac\Lambda3+\xi H^4\,,
\label{f1}
\eeq
\beq
\dot H +\frac32\,H^2=-\frac{\chi\,p}{2}+\frac{\Lambda}{2}+\xi H^2\at2\dot H+\frac32\,H^2\ct\,.
\label{f22}
\eeq
As usual $H=\dot a/a$ is the Hubble parameter. An equation similar to (\ref{f1}) has been obtained in a covariant renormalizable model for 
gravity \cite{sebacogno}. 

As a crucial consistent check, it is easy to show taht the latter equations give rise to matter continuity law 
in agreement with diffeomorphism invariance 
that is
\beq
\dot\rho=-3H(1+w)\rho\,,
\label{c123}
\eeq
 and also to the generalised Raychaudhuri equation
\beq
\ddot a=\dot H +H^2
=-\frac{\chi(1+3w)\,\rho}{6}+\frac{\Lambda}{3}+\xi H^2\at2\dot H+H^2\ct\,.
\label{f2}\eeq
In order to analyze the physical consequences, 
one may indifferently use two among the three equations (\ref{f1})-(\ref{c123}). 
An equation formally identical to (\ref{f1}) 
has already been obtained within AdS/CFT holographic correspondence in \cite{kiri,lidsey}, in brane cosmology \cite{marteens} 
or assuming Clausius relation and a logarithmic correction 
to the area law in \cite{cai1}.


The FLRW admits a dynamical trapping horizon (Hubble horizon) 
in the sense of Hayward \cite{sean} located at $R_H=\frac{1}{H}$,
with associated surface gravity and ''dynamical temperature'' \cite{us}
\beq
k_H=-\at H+\frac{\dot H}{2H}\ct\,,\hs\hs 
 T_H=\frac{\ka_H}{2\pi}=-\frac{1}{2\pi}\,\at H+\frac{\dot H}{2H}\ct\,.
\eeq
The identification of temperature with the suface gravity as it happens in the static case, 
is supported by a tunneling computation via 
the Hamilton-Jacobi method (see, for example, \cite{us,vanzorep} and references therein).

Now we introduce the generalised Misner-Sharp energy 
evaluated on the dynamical Hubble horizon, by means of
equation (\ref{c1}) obtained for the static case, then we get
\beq
E_H=\frac{1}{2G_N}\at R_H-\frac{\Lambda}{3}R_H^3-\frac{\xi}{R_H}\ct \,,
\eeq
and similarly for the Entropy,
\beq
S_H=\frac{A_H}{4G_N}-2\pi\xi\,\ln\frac{A_H}{4G_N}\,,
 \hs\hs A_H=4\pi R_H^2=\frac{4\pi}{H^2}\,.
\label{s1}
\eeq
Such quantities satisfy the Clausius relation as a consequence of field 
equations (\ref{f1})-(\ref{f2}).
In fact one has
\beq
d E_H=T_H dS_H \frac{ A_H}{4 G_N}-\frac{\chi}{16\pi}T_{(2)}dV_H\,.
\eeq
where $V_H=4\pi R_H^3/3$ is the volume of the Hubble sphere, while
$T_{(2)}=p-\rho=(w-1)\rho$ is the reduced stress tensor trace, 
a scalar quantity in a dynamical symmetric space time \cite{sean}. 
As for the static case, the cosmological constant is considered a given quantity. 


\section{Explicit solutions}

Here we assume $\rho\geq0$, $\La\geq0$ and analyze the behaviour of the solutions for
all possible values of $\xi$. For simplicity we also use units for which $\chi=1$. 

First of all, in the absence of matter, namely when $\rho(t)=0$,
there exist de Sitter solutions for which $\dot H_{dS}=0$. 
In fact, equations (\ref{f1}) and (\ref{f22}) have the trivial solutions
\beq
H^2_{dS}=\frac{1}{2\xi}\left(1\pm\sqrt{1-\frac{4\xi\Lambda}{3}}\right)\,,
\hs\hs\La>0\,,\xi>0\,,\hs \frac43\,\xi\La\leq1\,,
\eeq
\beq
H^2_{dS}=\frac{1}{2|\xi|}\left(-1+\sqrt{1+\frac{4|\xi|\Lambda}{3}}\right)\,,
\hs\hs\La>0\,,\xi<0\,.
\eeq
In the limit $|\xi\Lambda|\ll1$ one has (for expanding universe, $H>0$) 
\beq
\ag\begin{array}{lll}
H_1\simeq\sqrt{\frac{\Lambda}{3}}\,,&\hs\La>0\,,&\xi\geq0\,,\\\\
    H_2\simeq\sqrt{\frac{1}{\xi}}\,.&\hs\La\geq0\,,&\xi>0\,,\\\\
        H_3\simeq\sqrt{\frac{\Lambda}{3}}\,,&\hs\La>0\,,&\xi\leq0\,.
\end{array}\cp
\label{dSdS}\eeq
Both the solutions with $H(t)=H_1=H_3$ could describe current accerelation era independently on
$\xi$, while the solution with $H(t)=H_2$ could describe inflationary era independently on $\La$. 
It has to be stressed that in the case $\xi>0$, the model effectively 
describes only one of the two possible phases, because they correspond to distinct solutions.

Now we go back to the general model with arbitrary matter.
We first observe that in some cases from the generalised Friedmann equation
one gets constraints on the possible values of  $H$ and $\rho$ and, as a consequence,  
the corresponding solution is not singular for any finite value of $t$.
In fact, solving (\ref{f1}) for $H^2$ we obtain
\beq
H^2=\frac{1}{2\xi}\,\at1\pm\sqrt{1-\frac{4\xi(\rho+\La)}3}\ct\,,\hs\hs\xi>0\,,
\label{gamma}\eeq
\beq
H^2=\frac{1}{2|\xi|}\,\at-1+\sqrt{1+\frac{4|\xi|(\rho+\La)}3}\ct\,,\hs\hs\xi<0\,.
\label{gamma2}\eeq
If $\xi>0$, from (\ref{gamma}) we get the constraints
\beq
\frac{3}{4\xi}\geq\rho+\La\geq0\segue
\ag\begin{array}{l} 
\frac1{2}\geq \xi H^2\geq0\,,\\
\frac{1}{2}\leq \xi H^2\leq1\,,
\end{array}\cp\hs\hs\xi>0\,.
\label{CST}\eeq
If $\xi<0$ equation (\ref{gamma2}) does not give any constraint on 
$H$ and $\rho$. 

Since we would like to describe possible changes of phase
during the expansion, for example the exit from inflation or the beginning of
actual acceleration, we search for the stationary points of the quantity 
$\dot a(t)$. 
By solving the system of equations (\ref{f1}), (\ref{f2}) for $\dot H$ we get
\beq
\ddot a=H^2-\frac{(1+w)}{2(1-2\xi H^2)}\,\rho
       =-\frac{1}{2(1-2\xi H^2)}\,\aq (1-3w)\xi H^4+(1+3w)H^2-(1+w)\La\cq\,,
\label{dotH}\eeq 
and so $\ddot a(t)$ vanishes  when $H^2(t)$ reaches the (positive) values 
\beq
H^2_0=\ag\begin{array}{lll}
\frac23\,\La\,,&\hs \xi\neq0\,,&w=\frac13\,,\\ \\
 \frac{1+3w}{2\xi(1-3w)}\aq\,-1+\sqrt{1+\frac{4\xi\La(1+w)(1-3w)}{(1+3w)^2}}\,\cq\,,
       &\hs \xi>0\,,&w\neq1/3\,,\\ \\
   \frac{1+3w}{2|\xi|(1-3w)}\aq\,1\pm\sqrt{1-\frac{4|\xi|\La(1+w)(1-3w)}{(1+3w)^2}}\,\cq\,,
        &\hs \xi<0\,,&w\neq1/3\,.
\end{array}\cp
\label{H02}\eeq
This means that when $H^2(t_0)=H^2_0>0$ the system goes 
from a decelerated to an accelerated expansion (or viceversa), and this happens only if $H_0$ is in the permitted range of $H$.
If $H_0=0$, then there is a singularity ($a(t_0)=\ii$) or a bounce ($\dot a(t_0)=0$,
when $t_0<\ii$). 

In the case $\xi>0$ we must take into account of (\ref{CST}) and so,
in order to have a possible change of phase,
the free parameters have to satisfy the following constraints (we consider matter or radiation contributes),
\beq
\begin{array}{llll}
(\La>0\,,\xi>0\,,0\,,w=\frac13)\,, &\hs H_0^2=\frac23\,\La&\segue&
           \ag\begin{array}{l}0<\frac43\,\xi\La\leq1\,,\\
             1\leq\frac43\,\xi\La\leq2\,,\end{array}\cp
\\\\
(\La>0\,,\xi>0\,,w=0)\,, &\hs H_0^2=\frac{1}{2\xi}\at-1+\sqrt{1+4\xi\La}\ct&\segue&
           \ag\begin{array}{l}0<\frac43\,\xi\La\leq1\,,\\
                      1\leq\frac43\,\xi\La\leq\frac83\,.
\end{array}\cp
\end{array}
\eeq  
When $\La=0$, the only stationary point of $\dot a(t)$ corresponds to $H_0=0$ 
and so the model with $(\La=0,\xi>0)$ can describe only one phase of the expansion.

In the case $\xi<0$, the algebraic equation (\ref{gamma2}) does not give any restiction on
the values of $H^2$, but nevertheless, as we shall see in the following, the parameters 
$\xi$ and $\La$ are not completely arbitrary. Moreover,
from the third equation in (\ref{H02}) we also see that in this case, even for $\La=0$
 there is a non vanishing stationary point at $H_0^2=(1+3w)/|\xi|(1-3w)$ with $w\neq1/3$.

For the special case $\xi=0$ one can find
exact solutions for $a(t)$. In particular, the solutions of (\ref{f1})-(\ref{f22})
with $\rho>0$ read
\beq
a(t)=a_0\,\aq\sinh^2\,\at\frac{1+w}2\,\sqrt{3\La}\,t\ct\cq^\frac{1}{3(1+w)}\,,
\eeq
\beq
H(t)=\coth\,\at\frac{1+w}2\,\sqrt{3\La}\,t\ct\,,
\eeq
\beq
\rho(t)=\La\,\sinh^{-2}\,\at\frac{1+w}2\,\sqrt{3\La}\,t\ct\,,
\eeq
$a_0$ being an arbitrary constant.  
As one can see, $a(t)$ vanishes for a finite value of $t$ 
($t=0$ with the chosen initial conditions)
and as a consequence both  $H(t)$ and $\rho(t)$ are divergent at that time.

As we have seen above, in the absence of matter $H(t)=\La/3$ and 
$a(t)=a_0\,\exp(\sqrt{\La/3}\,t)$ (de Sitter solution).

If $\xi\neq0$ we are not able to get explicit solutions for $a(t)$, 
but nevertheless we can get implicit solutions for $H(t)$, 
which permits to understand the behaviour of the system. 

We start with the two cases $(\La=0,\xi>0)$ and $(\La=0,\xi<0)$. 
Apart arbitrary integration constants we have
\beq
\frac{3(1+w)}2\,t=\frac{1}{H}
      +\frac{\sqrt\xi}2\,\log\left|\frac{1+\sqrt\xi\,H}{1-\sqrt\xi\,H}\right|\,,
               \hs\hs\xi>0\,,
\label{xiGT0}\eeq
\beq
\frac{3(1+w)}2\,t=\frac{1}{H}
     -\sqrt{|\xi|}\,\arctan(\sqrt{|\xi|}\,H)\,,\hs\hs\xi<0\,.
\label{xiLT0}\eeq
It has to be noted that in order to have $\rho>0$, 
in (\ref{xiGT0}) $H(t)$ has to be restricted to the values 
$\xi H^2\leq1$ in agreement with (\ref{CST}).
This means that $a(t)$ does not vanish and the density does not diverge.
Equation (\ref{xiGT0}) effectively corresponds to two different solutions related to
the distinct algebraic equations in (\ref{gamma}). 
The  solution with $1/2\geq\xi H^2\geq0$ has an asymptotic behaviour of the kind 
$H(t)\sim 1/t\to0$, while the other with $1/2\leq\xi H^2\leq1$
goes as $H(t)\sim1/\sqrt\xi\,=\text{Const}$, giving rise to a de Sitter asymptotic behaviour 
for $a(t)$, that is
\beq
a(t)\sim a_0e^{t/\sqrt\xi}\,,\hs\hs \La=0\,,
     \hs \xi>0\,,\hs\frac{1}{2}\leq\xi H^2\leq1\,.
\eeq   
As we already said above, the model with such values of the free parameters 
describes only one phase of the expansion. It could be used in order to describe
the inflationary era as in Ref.~\cite{staro}, but it does not provide a 
natural exit from that phase.

In the second case, equation (\ref{xiLT0}), $H(t)$ has no restrictions  and so
there is a singularity for $t\to0$ ($a(t)\to0$ and $\rho(t)\to\ii$). 
This model is not able to describe inflation, because it does not have a
rapidly expanding phase, but it could describe current acceleration. 
In fact, putting for simplicity $w=0$ in (\ref{H02}), we obtain $H_0^2=1/|\xi|$ and
using (\ref{dotH}) we see that $\ddot a(t)$ is negative or positive 
according to whether $H(t)$ is smaller or greater that $H_0$.

Finally we consider the two general cases with $(\La>0,\xi>0)$ and $(\La>0,\xi<0)$.
Apart arbitrary integration constants, the implicit solutions for $H(t)$ are given by
\beq
\frac{3(1+w)}2\,t=
      \frac1{2\sqrt{\al_+}}\,\log\left|\frac{1+H/\sqrt{\al_+}}{1-H/\sqrt{\al_+}}\right|
      +\frac1{2\sqrt{\al_-}}\,\log\left|\frac{1+H/\sqrt{\al_-}}{1-H/\sqrt{\al_-}}\right|
         \hs\hs\La>0\,,\xi>0\,,
\label{LaxiGT0}\eeq
where
\beq
\al_\pm=\frac{1}{2\xi}\at1\pm\sqrt{1-\frac{4\xi\La}3}\ct\,,\hs\hs\xi\La\leq\frac34\,,
\eeq
and
\beq
\frac{3(1+w)}2\,t=
          \frac1{2\sqrt{\be_+}}\,\log\left|\frac{1+H/\sqrt{\be_+}}{1-H/\sqrt{\be_+}}\right|
          -\frac1{\sqrt{\be_-}}\,\arctan\at\frac{H}{\sqrt{\be_-}}\ct\,,
            \hs\hs\La>0\,,\xi<0\,,
\label{LaxiLT0}\eeq
where
\beq
\be_\pm=\frac{1}{2|\xi|}\at\sqrt{1+\frac{4|\xi|\La}3}\pm1\ct\,.
\eeq
Also in this case in order to have $\rho\geq0$, $H(t)$  in (\ref{LaxiGT0})
has to be restricted to the values $\al_-\leq H^2\leq\al_+$ and so $a(t)$ 
does not vanish for finite values of time.
The equation describes two distinct solutions, both of them giving rise to a
de Sitter asymptotic behaviour for $a(t)$ of the kind
\beq
a(t)&\sim&a_0 e^{\sqrt{\al_+}t}\,,\hs\lim_{t\to\ii}H(t)=\sqrt{\al_+}\,,
\label{atI}\\
a(t)&\sim&a_0 e^{\sqrt{\al_-}t}\,,\hs\lim_{t\to\ii}H(t)=\sqrt{\al_-}\,.
\eeq
Looking at (\ref{H02}) we see that  in both the cases $w=1/3$ and $w=0$ 
the stationary point is out of the permitted range
and so also this generalised solution can describe only one phase of the expansion.

In the last case, equation (\ref{LaxiLT0}), $\rho$ is non negative for $H^2\geq\be_-$
and so, in contrast with the analog case with $\La=0$, 
there is a the solution for which $a(t)$ is always finite for finite values
of time. From (\ref{LaxiLT0}) in fact we have 
\beq
\be_-\leq H^2(t)\leq\be_+\,,\hs\hs t_0\leq t\leq\ii\,,
\eeq 
\beq
\be_+\leq H^2(t)\leq\ii\,,\hs\hs \ii\geq t\geq t_0\,.
\eeq 
Both the latter solutions  
give rise to a de Sitter asymptotic behaviour for $a(t)$
similar to the one in (\ref{atI}), but with $\al_+$  replaced by $\be_+$.

Depending on the parameters, the previous solutions can describe decelerated and accelerated
expansion phases. First choosing $w=1/3$, from (\ref{H02}) we get 
\beq
H_0^2=\frac23\,\La\segue\ag\begin{array}{ll}
       \be_-<H_0^2<\be+\,&\mbox{ if }|\xi|\La<\frac94\,,\\
              H_0^2>\be+\,&\mbox{ if }|\xi|\La>\frac94\,,
\end{array}\cp\label{kaputt}
\eeq   
while in the case $w=0$ we obtain
\beq
H_0^2=H^2_\pm=\frac1{2|\xi|}\at1\pm\sqrt{1-4|\xi|\La}\ct
      \segue\ag\begin{array}{ll}
        \be_-<H^2_\pm<\be+\,&\mbox{ if }|\xi|\La<\frac14\,,\\
              H^2_\pm>\be+\,&\mbox{ never}\,.
\end{array}\cp\label{Oberfuhrer}
\eeq
Using the latter results now we are able to study the changes of phase of the system.
For example, if $w=1/3$ from (\ref{dotH}) it follows that 
$\ddot a$ is negative or positive according to whether $H^2(t)$ is higher or lower
than $H_0^2=2\La/3$. This means that the system passes from a decelerated
to an accelerated phase if $|\xi|\La<9/4$, while it passes from an accelerated to
a decelerated phase if $|\xi|\La>9/4$. Of course such last situation has to be
rejected for physical reasons, because both $\xi$ and $\La$ have to be small quantiies.  

In the case of $w=0$, the acceleration is negative for $H_-<H(t)<H_+$ and 
positive otherwise and this means that the model describes three phases, that is\\
1) accelerated expansion for $\be_-<H^2(t)<H^2_-$;\\
2) decelerated expansion for $H^2_-<H^2(t)<H^2_+$;\\
3) accelerated expansion for $H^2_+<H^2(t)<\be_+$.


\section{Concluding remarks}

In this paper, a toy model of Einstein gravity with a Gauss-Bonnet classically ``entropic'' term mimicking a quantum correction has been
 investigated. The static black hole solution due to Tomozawa has been recovered and generalized with the inclusion of non trivial 
horizon topology, and its entropy has been evaluated deriving the first law from equations of motion. As a result the Bekenstein-Hawking area 
law has acquired a corrected logarithmic area term. A Misner-Sharp expression for the mass of black hole has been found. 
The same model has been used in order to derive Friedmann equations with corrected terms, which reproduce the first law 
with the same formal entropy and energy of the static case, but now related to a general FLRW space-time, interpreted as a dynamical cosmological black hole, with a ``temperature `` associated with Hayward's dynamical surface gravity: this result is in agreement with one obtained by the tunneling method applied to FLRW space-time in Ref.~\cite{us}. A detailed  analysis of all possible  cosmological solutions including de Sitter and non singular solutions have been provided. These solutions describe several phases, the more interesting one is present for $\xi<0$ and $\Lambda >0$.
In this case, when $\omega=0$, there is no singularity at $t=0$, but  an initial power-like acceleration, followed by a deceleration phase, and by a final  exponential acceleration.

As a consequence, this model may be a candidate to describe inflation and current dark energy epoch with a natural exit and entrance in the various cosmological phases. 
Of course, it should be emphasized that in this qualitative work, 
we have focused only on the possibility 
of the realization of an unified scenario:
important issues of inflationary cosmology such as 
the  reheating process and the generation of the 
curvature perturbations with a power spectrum consistent with 
the anisotropies of the CMB, and the consistence with the observational constraints coming from observations of our universe~\cite{WMAP,ade}
are crucial in the choosing of boundary conditions and numerical tests on the model, which is not the aim of this work.

A generalization of this work may be done by using s-wave approximation and reduction to effective 2nd theory as in Ref.~\cite{ultimo}. It maybe also interesting to generalize the results of this work for $F(R)$-gravity (for review see Refs.~\cite{uno,due}, where number of BH solutions exist, as in the recent review~\cite{tre}).


\section*{Acknowledgments}
\paragraph*{}We thank 
Professor Sergei Odintsov 
for comments and valuable suggestions in this work.


{}

\begin{thebibliography}{}

\bibitem{tomo12} 
 Y.~Tomozawa,
 ``Quantum corrections to gravity,''
 arXiv:1107.1424 [gr-qc].

\bibitem{BD}
N D Birrell \& P C W Davies, Quantum fields in curved space
 (Cambridge University Press 1982).

\bibitem{smoot1} 
 D.~A.~Easson, P.~H.~Frampton and G.~F.~Smoot,
 Phys.\ Lett.\ B {\bf 696}, 273 (2011)
 [arXiv:1002.4278 [hep-th]].
\bibitem{smoot2} 
 D.~A.~Easson, P.~H.~Frampton and G.~F.~Smoot,
 Int.\ J.\ Mod.\ Phys.\ A {\bf 27}, 1250066 (2012)
 [arXiv:1003.1528 [hep-th]].

\bibitem{padma} 
  T.~Padmanabhan,
  Gen.\ Rel.\ Grav.\  {\bf 44}, 2681 (2012)
  [arXiv:1205.5683 [gr-qc]];
T.~Padmanabhan,
  ``Emergence and Expansion of Cosmic Space as due to the Quest for Holographic Equipartition,''
  arXiv:1206.4916 [hep-th].


\bibitem{staro} 
 A.~A.~Starobinsky,
 Phys.\ Lett.\ B {\bf 91}, 99 (1980).

\bibitem{duff} 
 M.~J.~Duff,
 Class.\ Quant.\ Grav.\ {\bf 11}, 1387 (1994)
 [hep-th/9308075];
 S.~Nojiri and S.~D.~Odintsov,
 Phys.\ Lett.\ B {\bf 444}, 92 (1998)
 [hep-th/9810008].

\bibitem{HS} 
 M.~Henningson and K.~Skenderis,
 JHEP {\bf 9807}, 023 (1998)
 [hep-th/9806087].

\bibitem{cai} 
 R.~-G.~Cai, L.~-M.~Cao and N.~Ohta,
 JHEP {\bf 1004}, 082 (2010)
 [arXiv:0911.4379 [hep-th]].

\bibitem{OdiBH}

S.~Nojiri and S.~D.~Odintsov,
 Int.\ J.\ Mod.\ Phys.\ A {\bf 15} (2000) 989
 [hep-th/9905089];
S.~Nojiri, O.~Obregon, S.~D.~Odintsov and K.~E.~Osetrin,
 Phys.\ Lett.\ B {\bf 458}, 19 (1999)
 [gr-qc/9904035].

\bibitem{Davood}
S.~H.~Hendi and D.~Momeni,
 Eur.\ Phys.\ J.\ C {\bf 71}, 1823 (2011)
 [arXiv:1201.0061 [gr-qc]].

\bibitem{son} 
 E.~J.~Son and W.~Kim,
 arXiv:1303.0491 [gr-qc].
 
\bibitem{seba} L.~Sebastiani and S.~Zerbini,
Eur.\ Phys.\ J.\ C {\bf 71}, 1591 (2011)
[arXiv:1012.5230 [gr-qc]]

\bibitem{supercai} 
R.~-G.~Cai,
  Phys.\ Rev.\ D {\bf 65}, 084014 (2002)
  [hep-th/0109133].

\bibitem{deser1}
 D.~G.~Boulware and S.~Deser,
 Phys.\ Rev.\ Lett.\ {\bf 55}, 2656 (1985).

\bibitem{h}
 J.~T.~Wheeler,
 Nucl.\ Phys.\ B {\bf 273}, 732 (1986).

\bibitem{KS} 
 A.~Kehagias and K.~Sfetsos,
 Phys.\ Lett.\ B {\bf 678}, 123 (2009)
 [arXiv:0905.0477 [hep-th]].
\bibitem{hora}
 P.~Horava,
 Phys.\ Rev.\ D {\bf 79}, 084008 (2009)
 [arXiv:0901.3775 [hep-th]].

\bibitem{haw} S. W. Hawking, Nature {\bf 248} 30 (1974);
Commun. Math. Phys. {\bf 43} 199-220 (1975).

\bibitem{PW}
M.~K.~Parikh and F.~Wilczek,
Phys.\ Rev.\ Lett.\ {\bf 85}, 5042 (2000).

\bibitem{ang} 
M. Angheben, M. Nadalini, L. Vanzo and S. Zerbini, JHEP {\bf 0505}, 014 (2005); 
M. Nadalini, L. Vanzo and S. Zerbini,
 J. Physics A Math. Gen. {\bf 39}, 6601 (2006).

\bibitem{gorbu}
G.~Cognola, O.~Gorbunova, L.~Sebastiani and S.~Zerbini,
Phys.\ Rev.\ D {\bf 84}, 023515 (2011)
[arXiv:1104.2814 [gr-qc]].

\bibitem{dima} 
 D.~V.~Fursaev,
 Phys.\ Rev.\ D {\bf 51}, 5352 (1995)
 [hep-th/9412161].

\bibitem{mann} 
 R.~B.~Mann and S.~N.~Solodukhin,
 Nucl.\ Phys.\ B {\bf 523}, 293 (1998)
 [hep-th/9709064].

\bibitem{kaul} 
 R.~K.~Kaul and P.~Majumdar,
 Phys.\ Rev.\ Lett.\ {\bf 84}, 5255 (2000)
 [gr-qc/0002040].

\bibitem{sen} 
 A.~Sen,
 arXiv:1205.0971 [hep-th].
\bibitem{bojo} 
 M.~Bojowald,
 arXiv:1209.3403 [gr-qc].

\bibitem{sebacogno} 
 G.~Cognola, E.~Elizalde, L.~Sebastiani and S.~Zerbini,
 Phys.\ Rev.\ D {\bf 83}, 063003 (2011)
 [arXiv:1007.4676 [hep-th]].

\bibitem{hu} 
 B.~Geyer, S.~D.~Odintsov and S.~Zerbini,
 Phys.\ Lett.\ B {\bf 460} (1999) 58
 [gr-qc/9905073];
S.~Nojiri, S.~D.~Odintsov and S.~Zerbini,
 Phys.\ Rev.\ D {\bf 62} (2000) 064006
 [hep-th/0001192];
S.~i.~Nojiri and S.~D.~Odintsov,
 Phys.\ Lett.\ B {\bf 484} (2000) 119
 [hep-th/0004097].
\bibitem{hu2} 
M.~V.~Fischetti, J.~B.~Hartle and B.~L.~Hu,
 Phys.\ Rev.\ D {\bf 20}, 1757 (1979).

\bibitem{vilenkin} 
 A.~Vilenkin,
 Phys.\ Rev.\ D {\bf 32}, 2511 (1985).

\bibitem{hertog} 
 S.~W.~Hawking, T.~Hertog and H.~S.~Reall,
 Phys.\ Rev.\ D {\bf 63}, 083504 (2001)
 [hep-th/0010232].
\bibitem{sean} 
 S.~A.~Hayward,
 Phys.\ Rev.\ D {\bf 49}, 831 (1994)
 [gr-qc/9303030]; S.~A.~Hayward,
 Phys.\ Rev.\ D {\bf 53}, 1938 (1996)
 [gr-qc/9408002]

\bibitem{us} 
 S.~A.~Hayward, R.~Di Criscienzo, L.~Vanzo, M.~Nadalini and S.~Zerbini,
 Class.\ Quant.\ Grav.\ {\bf 26}, 062001 (2009)
 [arXiv:0806.0014 [gr-qc]];
R.~Di Criscienzo, S.~A.~Hayward, M.~Nadalini, L.~Vanzo and S.~Zerbini,
 Class.\ Quant.\ Grav.\ {\bf 27}, 015006 (2010).

\bibitem{vanzorep} 
 L.~Vanzo, G.~Acquaviva and R.~Di Criscienzo,
 Class.\ Quant.\ Grav.\ {\bf 28}, 183001 (2011)
 [arXiv:1106.4153 [gr-qc]].
\bibitem{kiri} 
 F.~Bigazzi, R.~Casero, A.~L.~Cotrone, E.~Kiritsis and A.~Paredes,
 JHEP {\bf 0510}, 012 (2005)
 [hep-th/0505140].
\bibitem{lidsey} 
 J.~E.~Lidsey,
 Class.\ Quant.\ Grav.\ {\bf 26}, 147001 (2009)
 [arXiv:0812.2791 [gr-qc]].

\bibitem{marteens}
R. Maartens,
 Living\ Rev. Rel.\ {\bf 7} 7 (2004).


\bibitem{cai1} 
 R.~-G.~Cai, L.~-M.~Cao and Y.~-P.~Hu,
 JHEP {\bf 0808}, 090 (2008)
 [arXiv:0807.1232 [hep-th]].

\bibitem{WMAP}
E.~Komatsu {\it et al.}  [WMAP Collaboration],
  Astrophys.\ J.\ Suppl.\  {\bf 192}, 18 (2011)
  [arXiv:1001.4538 [astro-ph.CO]].

\bibitem{ade} 
  P.~A.~R.~Ade {\it et al.}  [Planck Collaboration],
  arXiv:1303.5076 [astro-ph.CO].

\bibitem{ultimo}
S.~Nojiri and S.~D.~Odintsov,
 Phys.\ Lett.\ B {\bf 463} 57 (1999).
 [hep-th/9904146].

\bibitem{uno}
S.~Nojiri and S.~D.~Odintsov,
  eConf C {\bf 0602061}, 06 (2006)
  [Int.\ J.\ Geom.\ Meth.\ Mod.\ Phys.\  {\bf 4}, 115 (2007)]
  [hep-th/0601213].

\bibitem{due}
S.~Nojiri and S.~D.~Odintsov,
  Phys.\ Rept.\  {\bf 505} (2011) 59
  [arXiv:1011.0544 [gr-qc]].

\bibitem{tre}
 R.~Myrzakulov, L.~Sebastiani and S.~Zerbini,
  ``Some aspects of generalized modified gravity models,'
  arXiv:1302.4646 [gr-qc] (2013).




\end{thebibliography}
\end{document}